\documentclass{elsart}

\usepackage{amssymb}

\begin{document}

\begin{frontmatter}



\title{Information Geometry and Phase Transitions}


\author[label1]{W. Janke},
\author[label2]{D.A. Johnston}
and 
\author[label3]{R. Kenna}
\address[label1]{Institut f\"ur Theoretische Physik, Universit\"at Leipzig,
Augustusplatz 10/11, 04109~Leipzig, Germany.}
\address[label2]{Department of Mathematics, Heriot-Watt University,
Riccarton, Edinburgh,  EH14 4AS, Scotland.}
\address[label3]{School of Mathematical and Information Sciences,
Coventry University,   Coventry, CV1 5FB, England.}

\begin{abstract}
The introduction of a metric onto the space of parameters
in models in Statistical Mechanics and beyond
gives an alternative perspective on their phase structure.
In such a geometrization, the scalar curvature, ${\mathcal{R}}$, 
plays a central role. A non-interacting model has a flat geometry
(${\mathcal{R}} = 0$),
while ${\mathcal{R}}$ diverges at the critical point of an interacting one.
Here, the information geometry is studied for 
a number of 
 solvable 
statistical-mechanical models.
\end{abstract}

\end{frontmatter}


\section{Introduction}
\label{Introduction}

As the application of statistical-physics techniques becomes more 
widespread and acceptable outside the traditional physics community,
it is clear that the phenomena of phase transitions have important 
roles to play there. Indeed, phase transitions are common to 
a very wide range of disciplines, from physics to biology, 
economics and even to sociology. As statistical physics finds 
pertinence in these areas, so too can the field draw on concepts 
outside the confines of pure physics.

In all of these disciplines, models are characterised by certain 
sets of parameters.
The idea of endowing the space of such parameters 
with a metric and geometrical structure
has been borrowed from parametric statistics \cite{Fish}.
Given a probability distribution $p(x|\theta)$, and a sample
$x_1, \dots, x_n$, the objective is to estimate the parameter 
$\theta$.
This may be done by maximizing the so-called 
likelihood function,
$
 L(\theta)
 =
 \prod_{i=1}^n{p(x_i|\theta)}
$,
or its logarithm (called the log-likelihood function),
\begin{equation}
 \ln{L(\theta)}
 =
 \sum_{i=1}^n{\ln{p(x_i|\theta)}}
\quad .
\end{equation}
The gradient of this quantity is the score function:
\begin{equation}
U(\theta)
=
\frac{d \ln{L(\theta)}}{d\theta}
\quad ,
\end{equation}
and the expectation of this random variable is zero
(${\rm{E}}[U(\theta)]=0$). Its variance is
\begin{equation}
{\rm{Var}}[U(\theta)]
=
{\rm{E}}
\left[
\frac{-d^2 \ln{L(\theta)}}{d\theta^2}
\right]
\quad .
\end{equation}
This quantity is called 
the expected or Fisher information.
Taylor expanding the log-likelihood function,
one arrives at
\begin{equation}
\ln{L(\theta + \delta \theta)}
-
\ln{L(\theta)}
=
\delta \theta
\left.
\frac{d \ln{L(\theta)}}{d\theta}
\right|_\theta
+
\frac{(\delta \theta)^2}{2}
\left.
\frac{d^2 \ln{L(\theta)}}{d\theta^2}
\right|_\theta
+ 
\dots
\quad .
\end{equation}
The first term on the right-hand side is zero at the true
$\theta$-value. Therefore,
the closeness of two probability distributions characterised
by $\theta$ and $\theta+\delta \theta$, is given by the
second term, or the
Fisher information.

For higher-dimensional distributions (which may be continuous), where, instead of $\theta$,
one has a set of parameters,
$\theta_1, \theta_2, \dots$, the Fisher information is 
defined as
\begin{equation}
 G_{ij} 
 (\theta)
 =
 -
 E
 \left[
 \frac{\partial^2 \ln{p(x|\theta)}}{\partial \theta_i \partial \theta_j}
 \right]
 =
 -\int{
 p(x|\theta)
  \frac{\partial^2 \ln{p(x|\theta)}}{\partial \theta_i \partial \theta_j}
dx
}
\quad .
\end{equation}
C.R.~Rao suggested this is a metric. It is, in fact, the
only suitable metric in parametric statistics and is called 
 the Fisher-Rao metric \cite{Fish}. 

In generic statistical-physics models, we have two parameters,
$\beta$, which we may think of as 
the inverse temperature, and $h$, the external field. In
this case the Fisher-Rao metric is simply given by 
\begin{equation}
G_{ij} = \partial_{i}\partial_{j} f \quad,
\label{frmetric}
\end{equation}
where $f$ is the reduced free energy per site and
$\partial_{i} = ( \partial/\partial\beta, \; \partial/\partial h)$.

For such a metric the scalar curvature may be calculated as
\begin{equation} 
{\mathcal R}\ =\ - \frac{1}{2 G^{2}} 
\left| \begin{array}{ccc} 
\partial^{2}_{\beta}     f & \partial_{\beta}\partial_{h}    f & 
\partial_{h}^{2}    f \\ 
\partial^{3}_{\beta}    f & \partial_{\beta}^{2}\partial_{h}    f & 
\partial_{\beta}\partial_{h}^{2}    f \\ 
\partial^{2}_{\beta}\partial_{h}    f & 
\partial_{\beta}\partial_{h}^{2}    f & \partial_{h}^{3}    f  
\end{array} \right| \quad , 
\label{equcurv} 
\end{equation}
where $G={\rm det}(G_{ij})$ is the determinant of the metric itself. 
The scalar curvature plays a central role in any attempt
to look at phase transitions from a geometrical perspective.
Indeed, ${\mathcal{R}}$ measures the complexity of the system.
A flat metric implies that the system is not interacting.
Conversely, and for all the models that have been considered
so far, the curvature diverges at (and only at) 
a phase transition point for
physical ranges of the parameter values.

Using standard scaling assumptions, we can anticipate the 
behaviour of ${\mathcal{R}}$ near a second-order critical point.
With $t=1-\beta/\beta_c$,
\begin{equation}
 f(\beta,h) 
 =
 \lambda^{-1} f \left( t\lambda^{a_t},h\lambda^{a_h} \right)
 =
 t^{\frac{1}{a_t}} 
 \psi
 \left( h t^{-\frac{a_h}{a_t}} \right)
\quad ,
\end{equation}
where
\begin{equation}
 a_t = \frac{1}{\nu d}
 \quad ,
 \quad \quad \quad
 a_h = \frac{\beta \delta}{\nu d}
\quad ,
\end{equation}
are the scaling dimensions for the energy and spin operators
and $d$ is the spacial dimensionality.
One finds for the scalar curvature,
\begin{equation}
{\mathcal R}\ =\ - \frac{1}{2 G^{2}} 
\left| \begin{array}{ccc} 
t^{\frac{1}{a_t} - 2} &  0                      &  t^{\frac{1}{a_t} -2\frac{a_h}{a_t}} \\ 
t^{\frac{1}{a_t} - 3} &  0                      &  t^{\frac{1}{a_t} -2\frac{a_h}{a_t} - 1}\\ 
0             &  t^{\frac{1}{a_t} -2\frac{a_h}{a_t} - 1}&  t^{\frac{1}{a_t} -3\frac{a_h}{a_t}}  
\end{array} \right| \quad , 
\label{reason}
\end{equation}
and 
\begin{equation}
G \sim t^{\frac{2}{a_t} +2\frac{a_h}{a_t} - 2}
\quad ,
\end{equation}
 yielding
\begin{equation}
 {\mathcal R}\ \sim \xi^d \sim |\beta - \beta_c|^{\alpha - 2}
\quad ,
\label{ex}
\end{equation}
where  hyperscaling 
($\nu d = 2 - \alpha $) is assumed and $\xi$ is the correlation length.

The systems hitherto analysed from the 
information-geometry perspective include
the Ising model in one dimension \cite{Jany}, 
the Bethe lattice Ising and mean-field models \cite{Brian}.
The main results hitherto established are
that ${\mathcal{R}}$ is positive definite
and 
diverges (as $\xi^d$ ) only at the 
critical point.

In an effort to see how generic these features are, and to discover
new ones, we analyse
 the Potts model in one dimension, 
the Ising model in two dimensions coupled to quantum gravity
and the spherical model in three dimensions.

\section{Information Geometry in Specific Models}
\label{Info}

{\bf{The Potts Model in One Dimension:}}
The one-dimensional $q$-state Potts model, like its Ising counterpart
(which corresponds to $q=2$) is exactly solvable.
Although it has no true phase transition, thermodynamic quantities 
diverge at zero temperature. 
In the Ising case, explicit calculations showed that
\begin{equation}
 {\mathcal{R}}_{\rm{Ising}} =
 1 + \frac{\cosh{h}}{\sqrt{\sinh^2{h} + \exp{(-4 \beta)}}}
\quad .
\end{equation}
The correlation length is known to behave as 
$
\xi 
\sim 
-1/\ln{(\tanh{\beta})}
$, and therefore
$
{\mathcal{R}}_{\rm{Ising}}
\sim 
\xi
$
as  $\beta \rightarrow \infty$.

\begin{figure}[htb]
\vspace{7cm}
\includegraphics{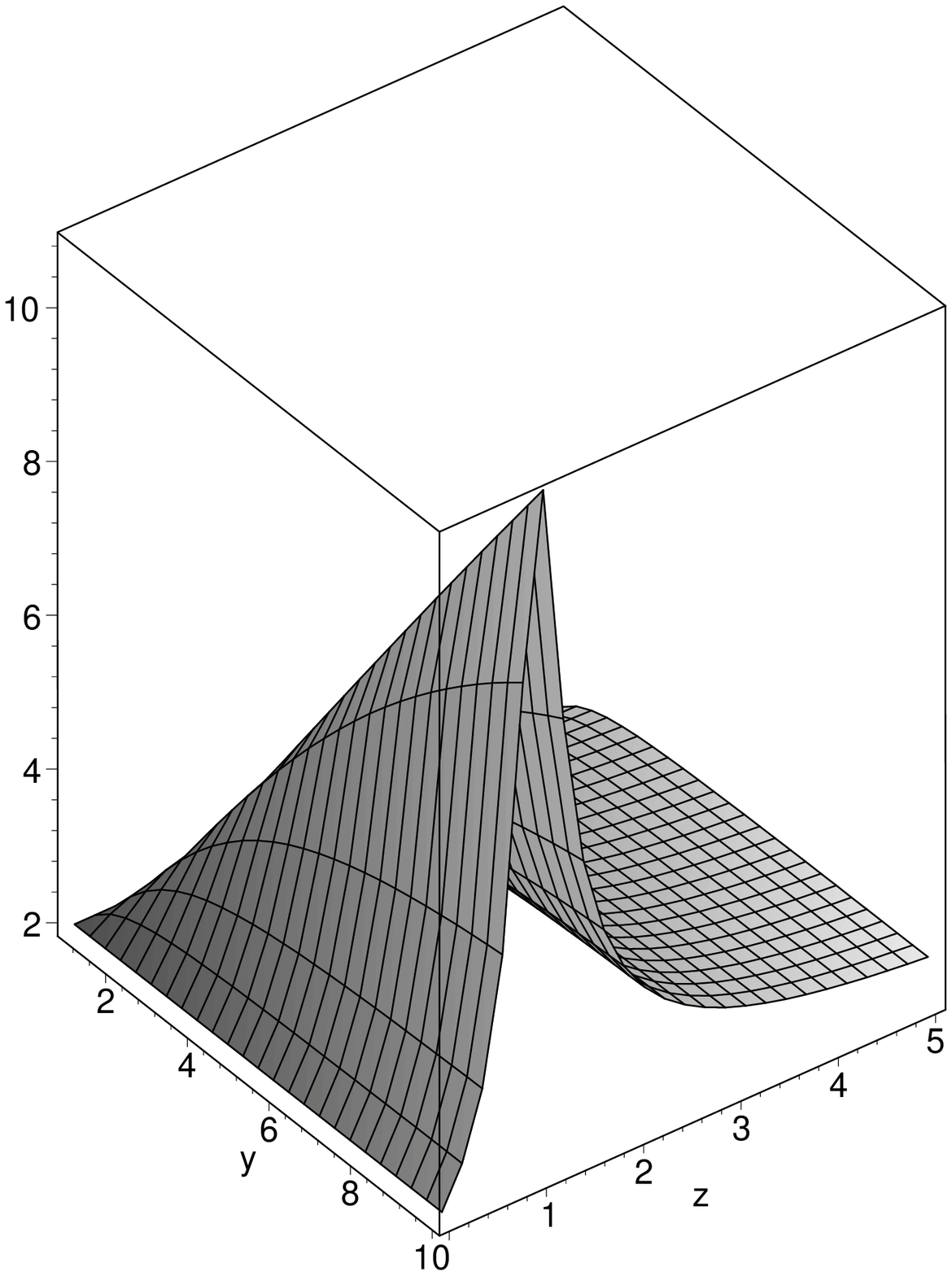}
\includegraphics{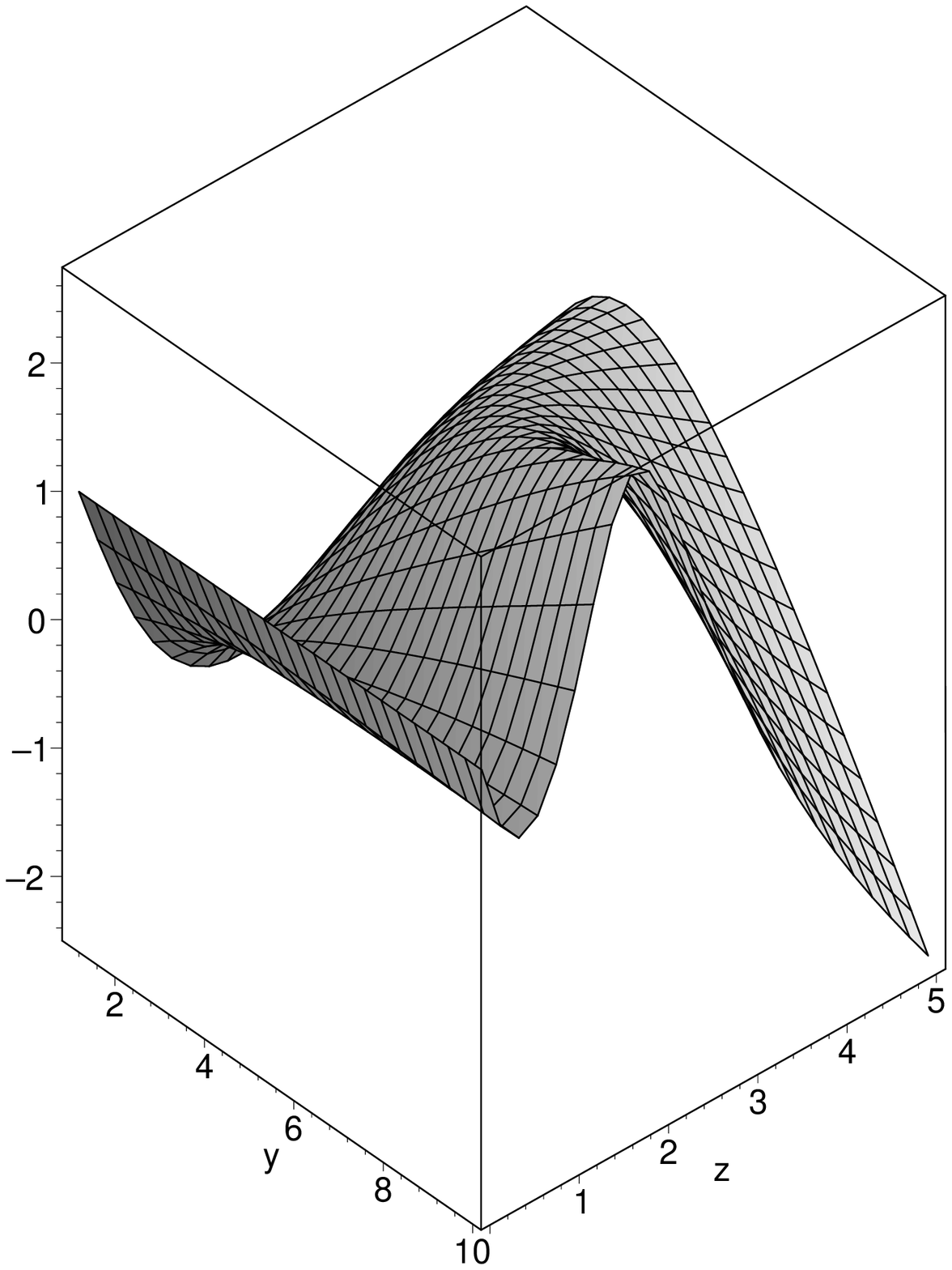}
\caption[a]{The scalar curvature for
the Ising model (left) and the 10-state Potts model
(right) in one dimension. }
\label{potts}
\end{figure}

For plotting purposes, it is convenient to introduce 
\begin{equation}
 y = \exp{(\beta)} \quad 
\quad \quad 
{\rm{and}}
\quad 
\quad \quad z = \exp{(h)}
 \quad.
\end{equation}
The curvature is plotted in Fig.~\ref{potts}.
It is clear that ${\mathcal{R}}_{\rm{Ising}}$ is positive definite and exhibits a 
$h \rightarrow -h$ (or $z \rightarrow 1/z$) symmetry.
Furthermore, given a particular value for
$y$, ${\mathcal{R}}_{\rm{Ising}}$ is maximum along
the zero-field ridge $z=1$.

The one-dimensional $q$-state 
Potts model is also exactly solvable, so
it is sensible to exploit it in an attempt to decide which 
of the previous features are generic. 
An explicit calculation \cite{us1} shows that
\begin{equation}
 {\mathcal{R}}_{\rm{Potts}}
 =
 A(q,y,z) + \frac{B(q,y,z)}{\sqrt{\eta(q,y,z)}}
 \sim
 y 
 \sim
 \xi
\quad ,
\end{equation}
for $h=0$ and $y \rightarrow \infty$.
Here,  $\eta(q,y,z)$ is the Potts
analogue of the Ising term, $\sinh^2{h} + \exp{(-4 \beta)}$,
and $A$ and $B$ are smooth functions of $y$ and $z$.
So the expected scaling (\ref{ex}) holds
with $d=1$ or $\alpha = 1$.

Figure~\ref{potts} also shows that, in the Potts case,
${\mathcal{R}}_{\rm{Potts}}$
is no longer positive definite. Furthermore, the $z \rightarrow
1/z$ symmetry of the Ising model is no longer present. 
For a given $y$, ${\mathcal{R}}_{\rm{Potts}}
$ no longer peaks along $z=1$.
Finally, an analysis of the Lee-Yang (complex $h$) 
zeroes of the model reveals that the curvature 
also diverges as the locus of zeroes, $z_0(y)$, is approached.
In fact,
\begin{equation}
{\mathcal{R}}
 \sim \left( z - z_0(y) \right)^{-\frac{1}{2}}
\quad ,
\end{equation}
so that this divergence is characterised by an edge exponent
$\sigma = 1/2$.

{\bf{The Three-Dimensional Spherical Model (and the
Ising Model on Planar Random Graphs):}}
The spherical model is given by
\begin{equation}
Z_{\rm{spherical}}
=
\int{
ds_1 \dots ds_{L^d}
\delta\left(\sum_is_i^2-L^d\right)
\exp{
    \left[
            \beta\sum_{\langle ij \rangle}{s_is_j}+ h\sum_is_i
    \right]
    }
}
\quad ,
\end{equation}
and can be solved by exponentiating the constraint and
using steepest descent. Allowing the  lattice extent, $L$,
to diverge reveals no transition for $d=1,2$.
The transition for $d=3$ has
$
\alpha = -1$,
$\beta = 1/2$
and
$\gamma = 2$.
This is the same set of critical exponents as for the 
Ising model on planar random graphs (matter coupled
to 2D gravity). This is remarkable, because there are no obvious
similarities between the two models.

Explicit calculations (in both models) give \cite{us2,us3}
\begin{equation}
 {\mathcal{R}} 
 \sim
 |\beta - \beta_c|^{-2}
\quad,
\label{jkb}
\end{equation}
which does {\emph{not}} accord with the prediction
of (\ref{ex}). That prediction is ${\mathcal{R}} 
 \sim
 |\beta - \beta_c|^{\alpha-2}
$, with $\alpha = -1$.
The source of the discrepancy can be traced back to the top left 
term in the determinant in (\ref{reason}), which is
$t^{-\alpha}$. In both current models, $\alpha$ is,
in fact, negative and this term  vanishes as criticality is approached. 
It is replaced by a constant term coming from the regular part of the
free energy. Both models then yield the same result (\ref{jkb}).

\section{Conclusions}
\label{Conclusions}

In statistical physics, and in related fields -- from the 
bio-sciences to economics -- phase transitions play a central role.
Thus new insights into the characterisation of critical phenomena
are of paramount importance. 
Here, geometric ideas from the field 
of parametric statistics are
``borrowed'' and explored. It is found that the curvature  associated
with the Fisher-Rao metric is, indeed, a useful quantity 
in the characterisation of phase transitions. 
Some features found in the Ising model are found not to be generic. 
In particular, ${\mathcal{R}}$ can be negative and there is no
 symmetry nor ridge along $h=0$. 
More surprisingly, 
the naively expected scaling behaviour
(\ref{ex}) fails in both the three-dimensional
spherical model and in the Ising
model on planar random graphs. Reasons for this are given.
Once again, it is curious that all critical exponents 
for the latter two models coincide, although there is no
obvious physical relation between them.

{\bf{Acknowledgements:}}
W.J. and D.J. were partially supported by
EC IHP network
``Discrete Random Geometries: From Solid State Physics to Quantum Gravity''
{\it HPRN-CT-1999-000161}.




\begin{thebibliography}{00}




\bibitem{Fish} R.A. Fisher, 
Phil. Trans. R. Soc. Lond. A {\bf 222} (1922) 309;\\
C.R. Rao, Bull. Calcutta Math. Soc. {\bf 37} (1945) 81.



\bibitem{Jany} 
 H. Janyszek, Rep. Math. Phys. {\bf 24}  (1986) 1; 
                                    {\it ibid.} 11;\\
 H. Janyszek and R. Mruga{\l}a, 
               Phys. Rev.  A    {\bf 39} (1989) 6515.


\bibitem{Brian}
  B.P. Dolan, Proc. Roy. Soc. London A {\bf  454} (1998) 2655.

\bibitem{us1} B.P. Dolan, D.A. Johnston, and R. Kenna, 
             J. Phys. A {\bf 35} (2002) 9025.

\bibitem{us2}  W. Janke, D.A. Johnston, 
and Ranasinghe P.K.C. Malmini,
Phys. Rev. E {\bf{66}} (2002) 056119.



\bibitem{us3}
W.~Janke, D.A.~Johnston, and R.~Kenna,
Phys. Rev. E {\bf{67}} (2003) 046106.

\end{thebibliography}
\end{document}